\documentclass[aps,preprint,superscriptaddress,showpacs]{revtex4}

\usepackage{epsfig,amsmath,amssymb}
\newcommand{\rv}{{\bf r}}
\newcommand{\xv}{{\bf x}}

\begin{document}

\title{Density functional for hard hyperspheres from a
  tensorial-diagrammatic series}

\author{Gavin Leithall}
\affiliation{H. H. Wills Physics Laboratory, University of Bristol, 
  Royal Fort, Tyndall Avenue, Bristol BS8 1TL, United Kingdom}
\author{Matthias Schmidt}
\affiliation{Theoretische Physik II,
  Universit\"at Bayreuth, D-95440 Bayreuth, Germany}
\affiliation{H. H. Wills Physics Laboratory, University of Bristol, 
  Royal Fort, Tyndall Avenue, Bristol BS8 1TL, United Kingdom}

\begin{abstract}
We represent the free energy functional by a diagrammatic series with
tensorial coefficients indexed by powers of length scale. For hard
cores, we obtain Percus' exact functional in one dimension and the
Kierlik-Rosinberg form of fundamental measures theory in three
dimensions. In five dimensions, the functional describes bulk fluids
better than Percus-Yevick theory does. At planar walls density
profiles oscillate with smaller periods than in lower dimensions. Our
findings open up avenues for treating both more general
high-dimensional systems, as well as three-dimensional mixtures via
dimensional reduction.
\end{abstract}


\date{30 July 2010, revised version: 21 October 2010, to appear in Phys. Rev. E (2011)}

\maketitle 

Studying the effects of inter-particle hard core repulsion has
provided significant insights into interfacial and capillary (phase)
behaviour of simple and complex liquids \cite{evans90}. Recent
examples include studies of ion-specific excluded-volume correlations
and solvation forces~\cite{kalcher10}, and the wetting properties of a
solid substrate by a ``civilized'' model of ionic
solutions~\cite{oleksy10}. Density functional theory (DFT)
\cite{evans79} is a primary tool for the investigation of such
situations. Applying DFT requires one to have an approximation for the
central (and in general unknown) Helmholtz free energy as a functional
of the one-body density distributions $\rho_i(\rv)$ of all species
$i$; $\rv$ is the spatial coordinate. One such example, which has
become a cornerstone of DFT, is Rosenfeld's fundamental-measure theory
(FMT) \cite{rosenfeld89,kierlik90,tarazona08review,roth10review} for
additive hard sphere mixtures.

There is considerable current interest in the study of the hard sphere
model in large space dimensions $D$. Motivation for such work
originates both from a desire to develop and to gain new insights into
the general structure of liquid state theories, as well as from the
question whether anything is special about $D=3$.  Studies that
exemplify this strategy include investigations of fluid structure
\cite{luban90,santos01,rohrmann07,gonzalesmelchor01} and of freezing
\cite{colotbaus,finken02dhs,vanMeel}.  Much less is known about
inhomogeneous hypersphere fluids.  In this Letter we present a
formalized framework for a generic hard hypersphere density
functional. Its structure is based on a density series, where the
central approximation lies in the star-like topology of the integral
kernels that couple the field points together
[cf.\ Eq.(\ref{EQhysSeries})].

The lowest order contribution to the exact virial expansion of the
excess (over ideal gas) free energy functional is $-k_BT\int d\rv_1
\rho_i(\rv_1) \int d\rv_2\rho_j(\rv_2) f_{ij}(|\rv_1-\rv_2|)/2$, where
the Mayer bond $f_{ij}(r)$ for hard spheres as a function of distance
$r$ is $f_{ij}(r)=-1$, if the two spheres with radii $R_i$ and $R_j$
overlap, and zero otherwise; $k_B$ is the Boltzmann constant and $T$
is temperature.  Graphically this can be represented by two filled
dots (each one corresponding to one multiplication by density and one
integration), that are joined by a line which stands for $f_{ij}(r)$.
The Kierlik-Rosinberg (KR) deconvolution of the Mayer bond for
arbitrary (odd) $D$ is $f_{ij}(|\rv_1-\rv_2|) = -\sum_{\mu=0}^D \int
d\xv w_\mu^i(|\rv_1-\xv|) w_{D-\mu}^j(|\rv_2-\xv|)$, with
species-dependent geometric ``weight'' functions $w_\mu^i(r)$; all
Greek indices run from 0 to $D$ here and in the following.  Hence
diagrammatically
\begin{align}
  -\frac{1}{2}\sum_{ij}
  \setlength{\unitlength}{0.5mm}
  \begin{picture}(50,20)(0,0)
    \put(4,2){\circle*{4}}
    \put(6,2){\line(1,0){19}}
    \put(22,7){$f_{ij}$}
    \put(4,9){$_i$}
    \put(42,9){$_j$}
    \put(25,2){\line(1,0){17}}
    \put(44,2){\circle*{4}}
  \end{picture}
  =
  \frac{1}{2}\sum_{ij}\sum_{\mu,\nu=0}^D
  \begin{picture}(50,20)(0,0)
    \put(4,2){\circle*{4}}
    \put(6,2){\line(1,0){17}}
    \put(4,9){$_i$}
    \put(42,9){$_j$}
    \put(11,-4){$_\mu$}
    \put(31,-4){$_\nu$}
    \put(20,8){$M^{\mu\nu}$}
    \put(21,0){$\times$}
    \put(26,2){\line(1,0){17}}
    \put(44,2){\circle*{4}}
  \end{picture}
  \equiv
  \frac{1}{2}
  \begin{picture}(50,20)(0,0)
    \put(4,2){\circle*{4}}
    \put(6,2){\line(1,0){17}}
    \put(21,7){$\sf M$}
    \put(21,0){$\times$}
    \put(26,2){\line(1,0){17}}
    \put(44,2){\circle*{4}}
  \end{picture},
  \label{EQdeconvolution}
\end{align}
where the cross indicates the position $\xv$ with integration and the
bonds labelled $\mu$ and $\nu$ represent $w_\mu^i(|\rv_1-\xv|)$ and
$w_\nu^j(|\rv_2-\xv|)$, respectively.  The second-rank tensor $\sf M$
has components $M^{\mu\nu}=M_{\mu\nu}=1$ for $\mu+\nu=D$ and is zero
otherwise \cite{schmidt07nage}.  The third diagram in
(\ref{EQdeconvolution}) is a short-hand notation, where e.g.\ the
first bond, together with its filled circle, represents $\sum_i \int
d\rv_1 \rho_i(\rv_1) w_\mu^i(|\rv_1-\xv|)\equiv n_\mu(\xv)$; tensor
contraction over all Greek indices is implied. Both $w_\mu^i$ and
$n_\mu$ possess units of (length)$^{\mu-D}$. In $D=3$, the
$n_\mu(\xv)$ are the familiar KR weighted densities \cite{kierlik90},
if the corresponding form of the $w_\mu^i(r)$ is used.

We approximate the exact ``triangle diagram'' of third order
\cite{footnote1} by a three-arm star:
\begin{align}
\setlength{\unitlength}{0.5mm}
-\frac{1}{6}
\sum_{ijk}
\begin{picture}(50,30)(0,0)
\put(6,-6){\circle*{4}}
\put(4,-14){$_i$}
\put(42,-13){$_j$}
\put(28,18){$_k$}
\put(5,8){$f_{ik}$}
\put(21,-14){$f_{ij}$}
\put(34,8){$f_{jk}$}
\put(6,-6){\line(3,4){18}}
\put(6,-6){\line(1,0){34}}
\put(42,-6){\line(-3,4){18}}
\put(42,-6){\circle*{4}}
\put(24,17){\circle*{4}}
\end{picture}
\approx
\frac{1}{6}
\sum_{ijk}\sum_{\mu,\nu,\tau=0}^D
\begin{picture}(50,30)(0,0)
\put(6,-6){\circle*{4}}
\put(4,-14){$_i$}
\put(42,-13){$_j$}
\put(28,18){$_k$}
\put(8,2){$_\mu$}
\put(30,-6){$_\nu$}
\put(18,10){$_\tau$}
\put(6,-6){\line(2,1){16}}
\put(30,7){$J^{\mu\nu\tau}$}
\put(21,0){$\times$}
\put(42,-6){\line(-2,1){16}}
\put(42,-6){\circle*{4}}
\put(24,18){\line(0,-1){14}}
\put(24,18){\circle*{4}}
\end{picture}
\equiv
\frac{1}{6}
\begin{picture}(50,30)(0,0)
\put(6,-6){\circle*{4}}
\put(6,-6){\line(2,1){16}}
\put(28,6){$\sf J$}
\put(21,0){$\times$}
\put(42,-6){\line(-2,1){16}}
\put(42,-6){\circle*{4}}
\put(24,18){\line(0,-1){14}}
\put(24,18){\circle*{4}}
\end{picture},
\end{align} \\ 
where the third-rank tensor $\sf J$ has components $J^{\mu\nu\tau}$,
which we assume i) to be symmetric under exchange of indices, ii) to
satisfy $J^{3\mu\nu}=M^{\mu\nu}$, and iii) to vanish if
$\mu+\nu+\tau\neq 2D$, because the diagram has to be of dimension
(length)$^0$.

Restricting ourselves to star topology, we assume the excess free
energy functional in $D$ dimensions to have the generic form
\begin{align}
\beta{\cal F}_{\rm exc}[\{\rho_i\}] =
\frac{1}{2} \setlength{\unitlength}{0.5mm}
\begin{picture}(50,30)(0,0)
\put(4,2){\circle*{4}}
\put(6,2){\line(1,0){17}}
\put(21,7){$\sf M$}
\put(21,0){$\times$}
\put(26,2){\line(1,0){17}}
\put(44,2){\circle*{4}}
\end{picture}
+\frac{1}{6}
\begin{picture}(50,30)(0,0)
\put(6,-6){\circle*{4}}
\put(6,-6){\line(2,1){16}}
\put(28,6){$\sf J$}
\put(21,0){$\times$}
\put(42,-6){\line(-2,1){16}}
\put(42,-6){\circle*{4}}
\put(24,18){\line(0,-1){14}}
\put(24,18){\circle*{4}}
\end{picture}
+\frac{1}{12}
\begin{picture}(50,30)(0,0)
\put(6,-6){\circle*{4}}
\put(6,-6){\line(2,1){16}}
\put(42,-6){\line(-2,1){16}}
\put(42,-6){\circle*{4}}
\put(21,8){$\sf JJ$}
\put(21,0){$\times$}
\put(7,10){\circle*{4}}
\put(7,10){\line(2,-1){16}}
\put(41,10){\line(-2,-1){16}}
\put(41,10){\circle*{4}}
\end{picture}
+\frac{1}{20}
\begin{picture}(50,30)(0,0)
\put(6,-6){\circle*{4}}
\put(6,-6){\line(2,1){16}}
\put(42,-6){\line(-2,1){16}}
\put(42,-6){\circle*{4}}
\put(18,-11){$\sf JJJ$}
\put(21,0){$\times$}
\put(7,10){\circle*{4}}
\put(7,10){\line(2,-1){16}}
\put(41,10){\line(-2,-1){16}}
\put(41,10){\circle*{4}}
\put(24,18){\line(0,-1){14}}
\put(24,18){\circle*{4}}
\end{picture} + \ldots,
\label{EQhysSeries}
\end{align}
\\
where $\beta=1/(k_BT)$, and
 the scalar coefficients are taken from the zero-dimensional
excess free energy
$\varphi_0(\eta)=(1-\eta)\ln(1-\eta)+\eta=\sum_{k=2}^\infty
\eta^k/[k(k-1)]=\eta^2/2+\eta^3/6+\eta^4/12+\cdots$; here $\eta$ is a
dummy variable.  The product $\sf JJ$ indicates tensor contraction via
$\sum_{\tau=0}^D J^{\mu\nu\tau}J_{\tau}^{\;\;\kappa\lambda}$, where
$J_\tau^{\;\;\kappa\lambda}=\sum_{\tau'=0}^D
M_{\tau\tau'}J^{\tau'\kappa\lambda} $; the triplet {\sf JJJ}
represents $\sum_{\tau,\sigma=0}^D
J^{\mu\nu\tau}J_\tau^{\;\;\kappa\sigma} J_\sigma^{\;\;\iota\zeta}$;
etc.
In order to find a closed expression for (\ref{EQhysSeries}), we
define a matrix of weighted densities
\begin{align}
  {\sf N}(\xv)&=
  \setlength{\unitlength}{0.5mm}
  \begin{picture}(15,10)(0,0)
    \put(8,0){$\sf J$}
    \put(0.8,-7){$\times$}
    \put(4,10){\line(0,-1){15}}
    \put(4,10){\circle*{4}}
  \end{picture},
  \label{EQtensorWeightedDensity}
\end{align}
where the spatial argument $\xv$ is not integrated over.  The
components of ${\sf N}({\bf x})$ are $N_\mu^{\;\;\nu}(\xv)=
\sum_i\sum_{\tau=0}^D J_\mu^{\;\;\nu\tau} (w_\tau^i \ast\rho_i)(\xv)$,
where $\ast$ denotes the $D$-dimensional convolution $(g\ast
h)(\xv)=\int d\rv g(\xv-\rv)h(\rv)$ of two functions $g(\xv)$ and
$h(\xv)$. The definition (\ref{EQtensorWeightedDensity}) allows one to
express the $k$-th order diagram in (\ref{EQhysSeries}) as a matrix
power ${\sf N}(\xv)^k$ with integration over $\xv$. Subsequently one
has to take the appropriate component $\mu=0,\nu=D$
\cite{footnote2}. Hence we can view (\ref{EQhysSeries}) as one
component of the tensorial functional
\begin{align}
  \beta {\sf F}_{\rm exc}[\{\rho_i\}] \equiv
  \int d\xv  \sum_{k=2}^\infty \frac{{\sf N}(\xv)^k}{k(k-1)}=
  \int d\xv \varphi_0({\sf N}(\xv)),
\end{align}
where the second equality comes from using the Taylor series of
$\varphi_0(\eta)$. The physically relevant functional ${\cal F}_{\rm
  exc}[\{\rho_i\}]$ is the $\mu=0,\nu=D$ component, i.e.\ ${\cal
  F}_{\rm exc}=({\sf F}_{\rm exc})_0^{\;\;D}$, which can be written as
$\beta{\cal F}_{\rm exc}[\{\rho_i\}]=\int d\xv\Phi$, where the free
energy density $\Phi=[\varphi_0({\sf N}(\xv))]_0^{\;\;D}$.

All properties specific to the value of $D$ are entirely encapsulated
i) in the dimensionality of the space integrals, ii) in the
prescription of the $D+1$ weight functions $w_\mu^i(r)$, and iii) in
the choice of the (constant) components $J^{\mu\nu\tau}$ of $\sf
J$. For all $D>0$, we use $w_D^i(r)=\Theta(r-R_i)$ and
$w_{D-1}^i(r)=\delta(r-R_i)$, where $\Theta(\cdot)$ is the Heaviside
(unit step) function and $\delta(\cdot)$ is the Dirac delta
distribution. For all cases considered, we could obtain the specific
form of the further weight functions and the form of $\sf J$ from the
minimalistic requirement that the two-body bulk fluid direct
correlation function $c_{ij}(|\rv-\rv'|)=-\delta^2 {\cal \beta F}_{\rm
  exc}/\delta \rho_i(\rv) \delta \rho_j(\rv')|_{\rho_k={\rm const}}$
is free of divergences, i.e.,
\begin{align}
  |c_{ij}(r)|<\infty, \quad {\rm for}\;{\rm all} \; r.
\end{align}

In $D=1$, $J^{011}=1$ determines $\sf J$ entirely and straightforward
algebra yields $\beta{\cal F}_{\rm exc}=-\int d\xv
n_0(\xv)\ln(1-n_1(\xv))$, which is exact
\cite{percus76,vanderlick89}. In $D=3$ we use the KR form for the
additional weight functions $w_1^i(r)$ and $w_0^i(r)$, and obtain
$J^{033}=J^{123}=1$, $J^{222}=1/(4\pi)$, which gives the KR functional
\cite{kierlik90}.  In $D=5$ we obtain
\begin{align}
  \Phi = -n_0\ln(1-n_5) + \frac{n_1n_4+n_2n_3}{1-n_5}
  +\frac{n_2n_4^2+n_3^2n_4/(64\pi^2)}{(1-n_5)^2}
  +\frac{n_3n_4^3}{48\pi^2(1-n_5)^3}
  +\frac{n_4^5}{160\pi^2(1-n_5)^4},
  \label{EQfreeEnergyDensity5D}
\end{align}
with the weight functions
$w_5^i=\Theta(R_i-r), \quad w_4^i=\delta(R_i-r)$,
$w_3^i=\delta'$,
$w_2^i=\delta'/(16\pi^2r)$,
$w_1^i=(8r^{-1}\delta''-\delta''')/(64\pi^2)$,
$w_0^i=(24r^{-3}\delta' + 24r^{-2}\delta''-12r^{-1}\delta'''+
    \delta'''')/(64\pi^2)$,
where $\delta'(R_i-r),\ldots,\delta''''(R_i-r)$ denote successive
derivatives of the Dirac delta function; the arguments have been left
away for clarity.

\begin{figure}
  \includegraphics[width=.7\columnwidth]{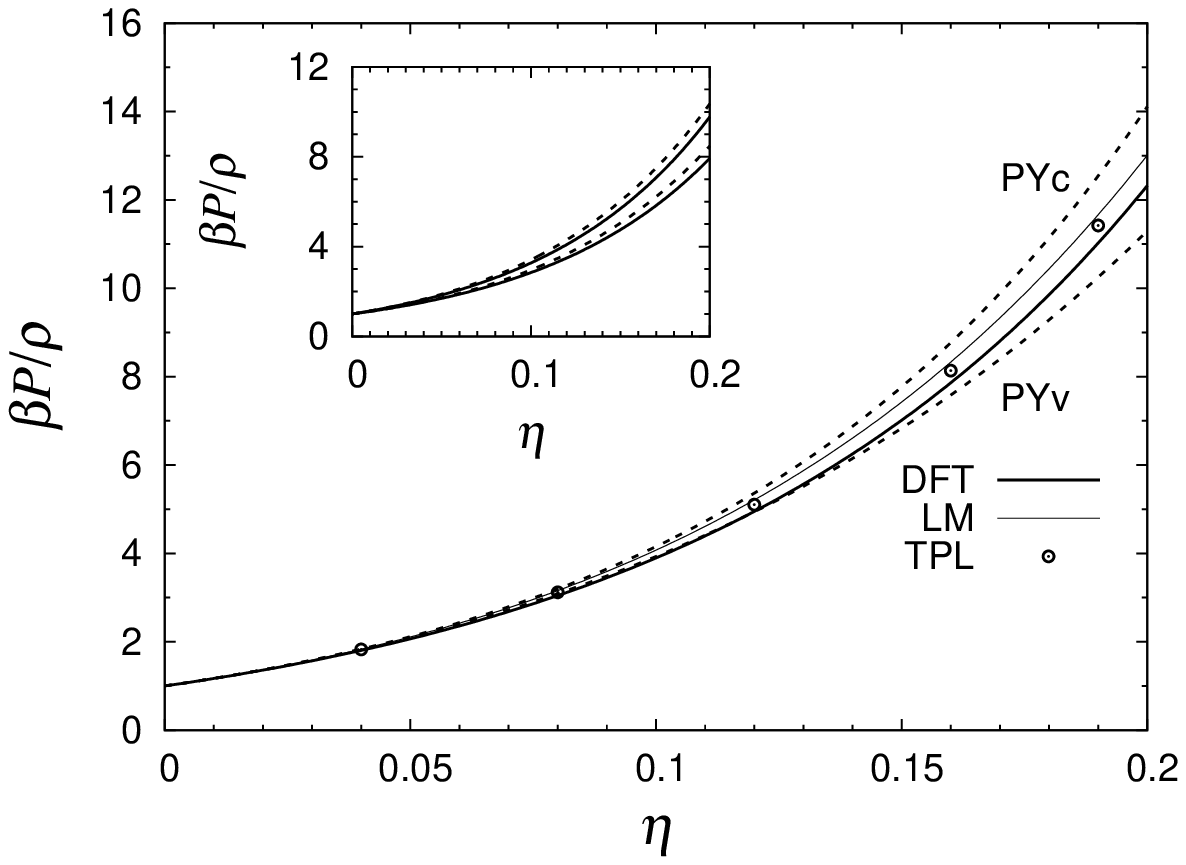}
  \caption{Compressibility factor $\beta P/\rho$ as a function of
    packing fraction $\eta$ for 5D hard hyperspheres, as obtained from
    the DFT bulk free energy (thick line), and compared to the PY
    compressibility (PYc) and virial (PYv) results (dashed lines), and
    the LM (thin line) equation of state \cite{luban90}. Also shown
    are DFT results from the test particle limit (TPL, symbols) using
    the virial theorem. The inset shows $\beta P/\rho$ as a function
    of the total packing fraction $\eta\equiv n_5$ for binary mixtures
    of 5D hard hyperspheres as obtained from the DFT (solid lines),
    along with the equation of state of Ref.~\cite{santos01} (dashed
    lines).  The upper pair of curves is for size ratio 4 and
    composition 0.75 of the larger species; the lower pair of curves
    is for an equimolar mixture with size ratio 2.5.}
  \label{fig1}
\end{figure}

For constant density fields the 5D functional yields an analytical
excess free energy for bulk fluids, $F_{\rm exc}(\{\rho_i\})={\cal
  F}_{\rm exc}[\{\rho_i={\rm const}\}]$.  When scaled as $\beta F_{\rm
  exc}/V$ this equals (\ref{EQfreeEnergyDensity5D}), but with
$n_\mu=\sum_i \xi_\mu^i \rho_i=\rm const$, where the fundamental
measures, $\xi_\mu^i=\int d\rv w_\mu^i(\rv)$, are $\xi_0^i=1$,
$\xi_1^i=R_i$, $\xi_2^i=R_i^2/2$, $\xi_3^i=32 \pi^2 R_i^3/3$,
$\xi_4^i=8 \pi^2 R_i^4/3$, and $\xi_5^i=8 \pi^2 R_i^5/15$.  The total
packing fraction is simply $\eta\equiv n_5$. The pressure is obtained
from $P=-\partial F_{\rm exc}/\partial V+k_BT \rho$, where
$\rho=\sum_i\rho_i$.  Fig.~\ref{fig1} shows $\beta P/\rho$ for pure 5D
hard hyperspheres as a function of $\eta$.  The DFT result lies
between the two Percus-Yevick (PY) equations of state; these deviate
more strongly from each other than in three dimensions (3D). Moreover,
the DFT only slightly underestimates, over the full range of (fluid)
packing fractions, the semi-empirical Luban-Michels (LM) equation of
state \cite{luban90}, which yields excellent agreement with simulation
data. For two representative binary mixtures (inset of
Fig.~\ref{fig1}), the DFT slightly underestimates the very accurate
equation of state of Ref.~\cite{santos01}, with similar (small)
deviations as in the pure case above.

\begin{figure}
  \includegraphics[width=.7\columnwidth]{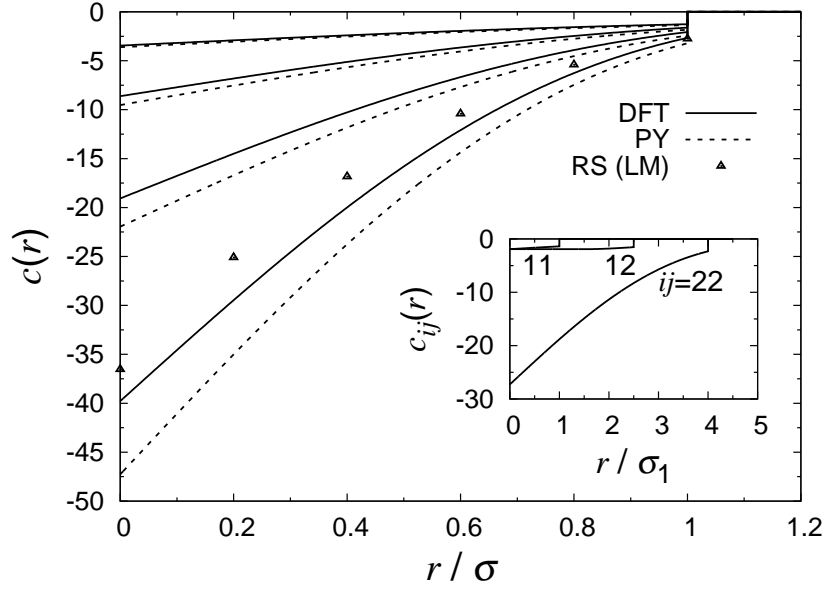}
  \caption{Bulk two-body direct correlation function, $c(r)$, as a
    function of the scaled distance $r/\sigma$, for 5D hard
    hyperspheres, as obtained from the DFT via second functional
    derivative (solid lines), along with the PY result (dashed lines),
    for packing fractions $\eta = 0.05, 0.1, 0.15, 0.2$ (from top to
    bottom). Also shown is the result of the theory by Rohrmann and
    Santos (RS) \cite{rohrmann07} (symbols), which is based on the LM
    equation of state \cite{luban90}, for $\eta = 0.2$ (we omit the
    small, positive Yukawa tail outside of the core). The inset shows
    DFT results for the partial pair direct correlation functions
    $c_{ij}(r)$ as a function of $r/\sigma_1$, where $\sigma_1=2R_1$,
    for a binary mixture of 5D hard hyperspheres with size ratio 4,
    composition of 0.75 for the (larger) species 2 and total density
    $\rho \sigma_1^5=1.4$, corresponding to $\eta = 0.173$ (cf.\ inset
    of Fig.~\ref{fig1}).  }
  \label{fig2}
\end{figure}

\begin{figure}
  \includegraphics[width=.7\columnwidth]{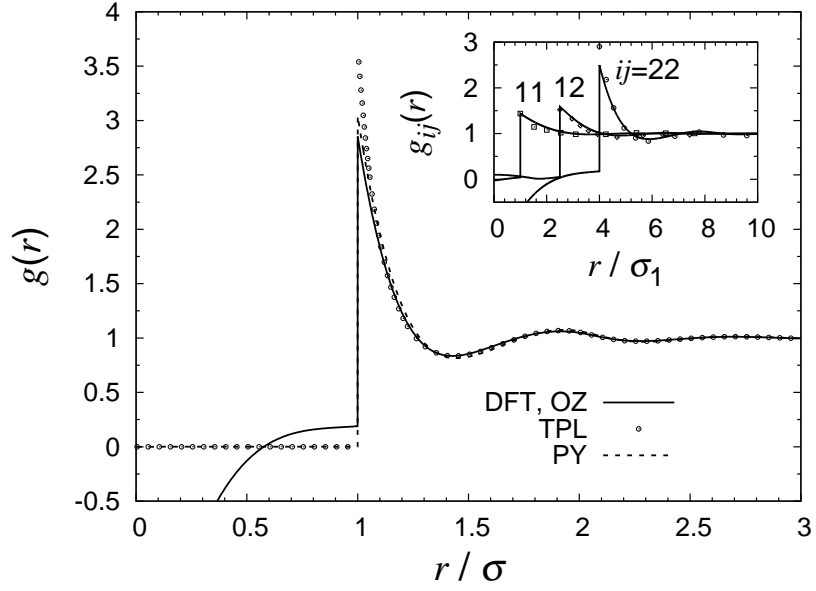}
  \caption{The radial distribution function, $g(r)$, as a function of
    $r/\sigma$, for 5D hard hyperspheres as obtained from the DFT
    using the OZ equation (solid line), using the test particle method
    (symbols), and compared to the PY result (dashed line) for packing
    fraction $\eta = 0.2$. The OZ result reaches $g(0)=-2.82$ at zero
    separation. The inset shows the partial pair distribution
    functions $g_{ij}(r)$ as obtained from the OZ route (lines) for
    the binary mixture described in the inset of Fig. \ref{fig2},
    along with simulation results \cite{gonzalesmelchor01} (symbols).}
  \label{fig3}
\end{figure}

Fig.~\ref{fig2} shows the pair direct correlation function $c(r)$ for
the pure 5D hard hypersphere fluid.  Outside of the core, the direct
correlation function vanishes, $c(r>\sigma)=0$, and it is a
fifth-order polynomials in $r$ for $r<\sigma$, similar to the PY
result; $\sigma=2R$ is the particle diameter.  However, the DFT result
is smaller in magnitude than in PY theory. This constitutes an
improvement, as can be observed by comparing to the sophisticated
theory by Rohrmann and Santos (RS) \cite{rohrmann07}, which is built
on the LM equation of state. For $\eta=0.2$, a value higher than that
at freezing \cite{vanMeel}, we find that indeed the DFT result
deviates much less from the RS theory than PY does. The inset of
Fig.~\ref{fig2} shows an example of partial pair direct correlation
functions, calculated from the functional. As expected, these satisfy
$c_{ij}(r>R_i+R_j)=0$ and they are free of divergences.
Fig.~\ref{fig3} displays corresponding results for the radial
distribution function $g(r)$, obtained from the Ornstein-Zernike (OZ)
relation with $c(r)$ as input. Outside of the core the deviation
between PY and DFT is very small, but the contact value from DFT is
lower and some core violation is apparent.  We also show results from
a test particle calculation, where we have numerically minimized the
density functional in the presence of an external potential that
equals the pair interaction potential.  Here the core condition is
automatically satisfied. The contact value $g(\sigma^+)$ increases,
and becomes larger than the PY value. When input into the virial
theorem, $\beta P/\rho=1+2^{D-1}\eta g(\sigma^+)$, the result is
almost indistiguishable, see Fig.~\ref{fig1}, from the quasi-exact LM
result.  The inset of Fig.~\ref{fig3} shows the partial pair
distribution functions $g_{ij}(r)$ for the mixture considered above
(inset of Fig.~\ref{fig2}). The agreement with simulation results
\cite{gonzalesmelchor01} is excellent.  Hence the DFT gives very good
account of both the thermodynamics and the structure of pure and
binary bulk liquids of 5D hard hyperspheres.

\begin{figure}
  \includegraphics[width=.7\columnwidth]{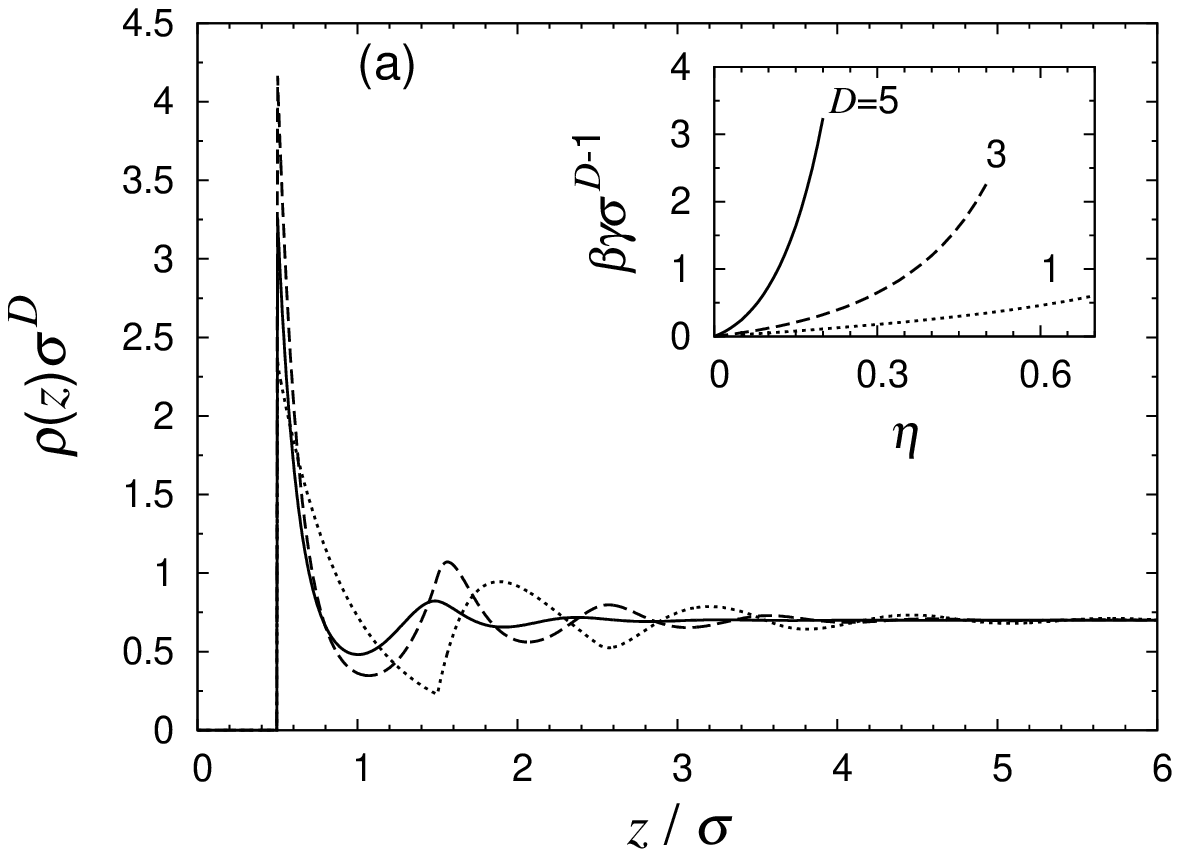}
  \includegraphics[width=.7\columnwidth]{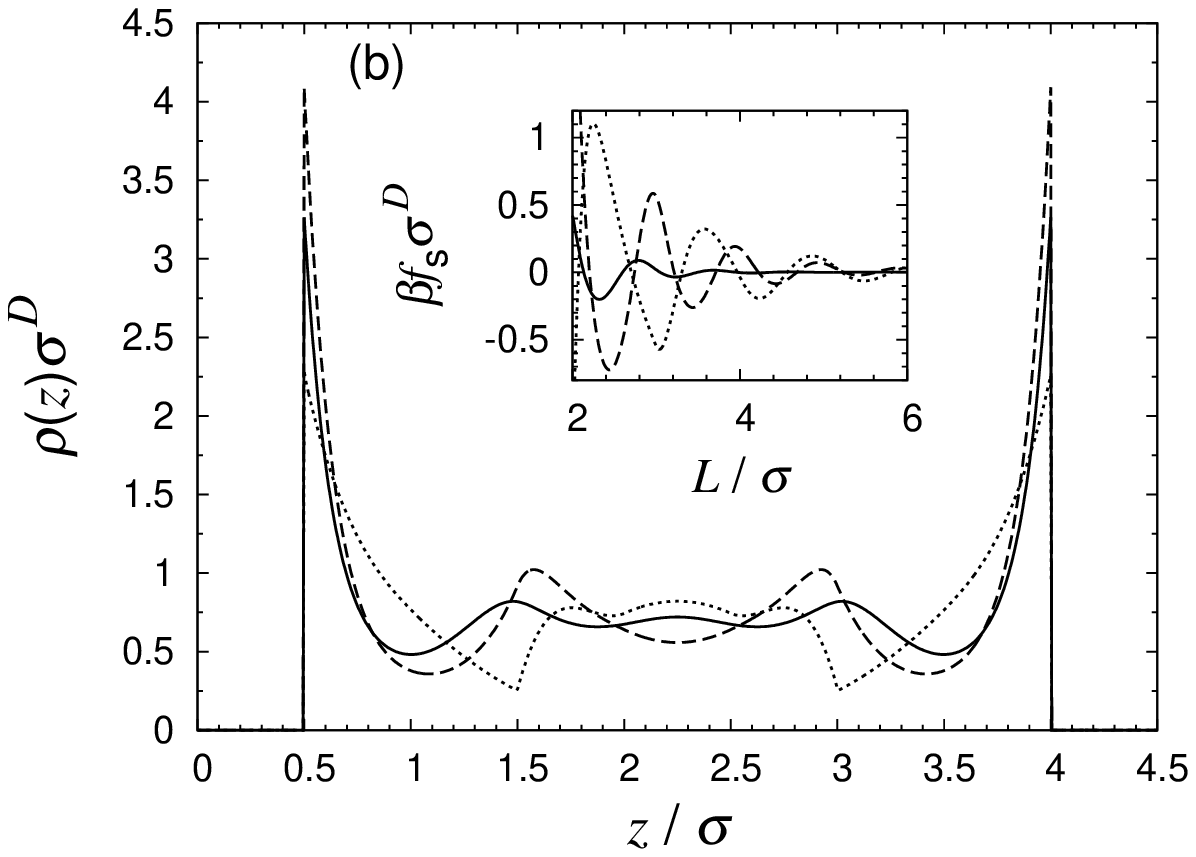}
  \caption{Scaled density profiles $\rho(z)\sigma^D$ as a function of
    the scaled distance $z/\sigma$ for packing fraction $\eta=0.7$ in
    1D (dotted line), 0.367 in 3D (dashed line) and 0.115 in 5D (solid
    line).  (a) Hard hypersphere fluids adsorbed at a hard wall; $z$
    is the distance from the wall. The inset shows the surface tension
    $\beta \gamma \sigma^{D-1}$ due to the presence of the wall as a
    function of $\eta$. (b) Density distributions for hard hypersphere
    fluids in a planar slit with width $L=4.5\sigma$ in chemical
    equilibrium with a bulk having the same value of $\eta$ as given
    above. The inset shows the scaled solvation force $\beta f_{\rm
      S}\sigma^D$ as a function of $L/\sigma$.}
  \label{fig4}
\end{figure}

Density profiles at a hard wall are shown in Fig.~\ref{fig4}(a) for
the cases 1D, 3D and 5D, keeping the scaled bulk density
$\rho\sigma^D=0.7=\rm const$.  With increasing dimensionality, the
distance between the liquid layers that are induced by the wall,
decreases strongly. The surface tension $\gamma=(\Omega+pV)/A$, where
$\Omega$ is the grand potential, $A$ is the area of the wall, and $V$
is the system volume $z\geq 0$, is shown in the inset of
Fig.~\ref{fig4}(a) \cite{footnote3}. This is seen to rise much more
strongly with $\eta$ for increasing $D$.  Fig.~\ref{fig4}(b) shows
density profiles in a slit for 1D (hard rods on a line), 3D, and for
5D. Again very different structuring is apparent, with the position of
the secondary peaks being closest to the wall for 5D and five (four)
layers present in 5D (3D).  The inset of Fig.~\ref{fig4}(b) shows the
solvation force $\beta f_{\rm S}(L)$, which for hard walls can be
obtained from $\beta f_{\rm S}(L)=\rho(\sigma^+/2)-\beta P$, as a
function of the slit width $L$. This again highlights sensitive
dependence on dimensionality.

Our formalized diagrammatic approach to DFT has significant relevance
for future developments: i) A whole class of FMT functionals for 3D
models can be generalized in a straightforward way to high
dimensions. This includes the penetrable sphere model of particles
interacting with a repulsive step function pair potential
\cite{schmidt99ps} (via adapting the scalar coefficients in
Eq.\ (\ref{EQhysSeries}) in order to allow for thermally induced
overlap of particles), the Asakura-Oosawa colloid-polymer mixture
\cite{schmidt00cip} (via linearizing the hypersphere mixtures
functional in one of the densities, which then describes an ideal
polymer species), or the Widom-Rowlinson model \cite{schmidt01wr} (by
using the appropriate expression for the zero-dimensional free energy
in order to generate the scalar coefficient in
Eq.~(\ref{EQhysSeries})). ii) Novel functionals for 3D models can be
obtained via dimensional {\em reduction}, where the hard hypersphere
functional (or one of its descendants above) is exposed to dimensional
crossover, such that the density distribution is assumed to be a delta
function (or superposition several thereof) in one or more space
dimensions. The particles in each hyperspace then constitute an
individual species in 3D. Mixtures with negative non-additivity can be
obtained via this procedure, see e.g.\ Ref.\ \cite{RSLTlong} for a
corresponding example of crossover from 3D to 2D.  iii) Although we
have used the KR approach, very interesting and useful further
insights could be gained from generalizing Rosenfeld's more geometric
formulation \cite{rosenfeld89} to high dimensions. This also applies
to its refinements based on considering zero-dimensional cavities
\cite{tarazona97,tarazona00}, possibly along the initial steps taken
in Ref.\ \cite{finken02dhs}.

We thank R.\ Evans and P.\ Hopkins for useful discussions. This work
was supported by the EPSRC under Grant No.\ EP/E065619 and by the DFG
via SFB840/A3.

\clearpage

\begin{thebibliography}{10}
\bibitem{footnote1} The rapid convergence of the virial series with
  increasing dimensionality is true only for homogeneous (bulk) fluid
  states.
\bibitem{footnote3} We have numerically tested that $\beta
  P=\rho(\sigma^+/2)$ and $-\partial \gamma/\partial \tilde\mu = \int
  dz [\rho(z)-\rho(\infty)]$, where $\tilde \mu$ is the chemical
  potential, are satisfied.  For the slit we have checked $f_{\rm S} =
  -2(\partial \gamma/\partial L)_{\tilde \mu,A}$.
\end{thebibliography}
\end{document}